**An experimental design for the control and assembly of magnetic microwheels**

E.J. Roth,[1] C.J. Zimmermann,[2] D. Disharoon,[2] T.O. Tasci,[3] D.W.M. Marr,[2] K.B. Neeves[1,4]

[1]Department of Bioengineering, University of Colorado Denver | Anschutz Medical Campus

[2]Department of Chemical and Biological Engineering, Colorado School of Mines

[3]Center for Bioengineering in Medicine, BioMEMS Resource Center, Harvard Medical School

[4]Department of Pediatrics, University of Colorado Denver | Anschutz Medical Campus

# ABSTRACT

Superparamagnetic colloidal particles can be reversibly assembled into wheel-like structures called microwheels (μwheels) which roll on surfaces due to friction and can be driven at user-controlled speeds and directions using rotating magnetic fields. Here, we describe the hardware and software to create and control the magnetic fields that assemble and direct μwheel motion and the optics to visualize them. Motivated by portability, adaptability and low-cost, an extruded aluminum heat dissipating frame incorporating open optics and audio speaker coils outfitted with high magnetic permeability cores was constructed. Open-source software was developed to define the magnitude, frequency, and orientation of the magnetic field, allowing for real time joystick control of μwheels through two-dimensional (2D) and three-dimensional (3D) fluidic environments. With this combination of hardware and software, μwheels translate at speeds up to 50 μm/s through sample sizes up to 5 cm x 5 cm x 5 cm using 0.75-2.5 mT magnetic fields with rotation frequencies of 5-40 Hz. Heat dissipation by aluminum coil clamps maintained sample temperatures within 3 ˚C of ambient temperature, a range conducive for biological applications. With this design, μwheels can be manipulated and imaged in 2D and 3D networks at length scales of micrometers to centimeters.

## I. INTRODUCTION

Microscale bots have potential for targeted cargo delivery in biological systems [1-3]; however, motile efficiency of microbots in complex environments is a challenge. Given their small size, these microbots require propulsion strategies that can overcome the constraints of low Reynolds number fluid mechanics where viscous forces dominate and translation is only possible by symmetry breaking [4, 5]. Successful strategies include bio-inspired swimmers with helical features or flexible oars [6, 7], catalytic particles that use chemical reactions to create thrust [5], and rod-like walkers, wheels, and lassos [3, 8, 9] that break symmetry by their interaction with surfaces. To assemble, power, and direct such microbots, externally applied magnetic fields [9-11] can be used and here we describe an apparatus for magnetic field control of rolling microbots called µwheels [12].

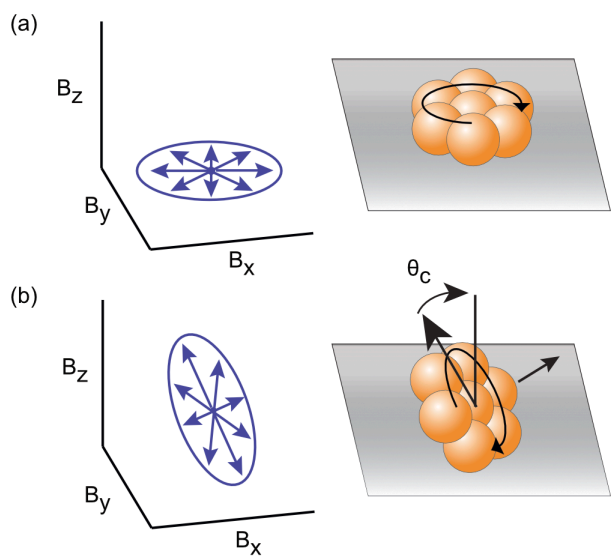

**FIG. 1.** (a) A $B_x$-$B_y$ magnetic field acting only in the x-y plane causes µwheels to sit and spin. (b) Adding a $B_z$ component causes µwheels to stand and roll. The camber angle, $\theta_c$, defines the µwheel tilt relative to the normal.

µWheels are assembled bottom-up from superparamagnetic colloids that, upon application of a rotating magnetic field, form close-packed, spinning aggregates due to attractive induced dipole interactions (Fig. 1). They propel themselves by rolling via friction along available surfaces [12]. Their speed is controlled by the magnetic field rotation frequency with maximum velocities determined by field magnitude and slip at the wall. Speed can be enhanced by applying external loads via DC magnetic fields that draw the assemblies closer to the surface thereby reducing slip [13]. µWheels are readily steered in any direction by changing the orientation of the magnetic field used to assemble and rotate them (Fig. 1). µWheels functionalized with biological molecules can do both biochemical and biomechanical work, for example in the ablation of blood clots using immobilized fibrinolytic enzymes [14]. Since the individual particles from are superparamagnetic, µwheels disassemble when the magnetic field is turned off. As such, µwheels can be assembled, directed to perform a task, and then disassembled for ready removal in confined geometries.

To use µwheels in complex environments such as the mammalian vasculature, it is necessary to design an apparatus to develop and test µwheel driving strategies in tortuous three-dimensional environments. The apparatus described in this article builds upon our prior work [12, 13], with improvements in imaging, heat dissipation and control software. The imaging stage has improved capabilities including a 10 cm × 10 cm × 10 cm test section large enough to accommodate a mouse. The design includes an extruded aluminum heat dissipating frame, open optics, and speaker coils with high magnetic permeability cores. The apparatus provides a magnetic field strength of 2.5 mT at the center of the sample, enabling µwheels to travel up to 50 µm/s while keeping sample temperature within a few degrees Celsius of ambient temperature. Removeable ferrite plugs placed in the center of the coils can increase the field strength in a 3 cm × 3 cm × 3 cm test section. Open source software was developed and used to define the magnitude, frequency, and orientation of the magnetic fields. The apparatus was tested by driving µwheels on a planar surface and through a helical channel.

**II. METHODS**

**A. Apparatus Summary**

To create suitable magnetic field strengths over larger samples, field amplitudes were increased using designs that included ferrite cores within copper wound coils with higher magnetic permeability than the air-cores used in prior work [3, 12-14]. To allow for positioning of large samples within the test section, a sample holder with three-dimensional translation capabilities was built. Two interchangeable microscopes were designed, one for imaging within large samples using an f-theta lens with a long working distance and the other using short focal distance objectives for high resolution images on planar samples. Resistive heating in the coils can cause substantial heating of samples and reduce the continuous operation time of the apparatus, and thereby limiting the amount of current that can be used. Heat dissipation components were therefore integrated into the coil holders and apparatus frame to reduce sample heating and prolong experiment run time. The apparatus was designed with a modular rail system made from extruded aluminum which can be modified in a piecewise fashion as experimental needs change using easily sourced, relatively low-cost materials (Table S1). An open-source software package used to create the waveforms necessary to drive the μwheels was developed and is available online [15].

**B. Apparatus Frame**

A custom frame with an open architecture was built to house the apparatus components (Fig. 2a). The 3 ft wide × 2 ft deep × 4.5 ft tall frame was built from modular extruded aluminum sections (Tnutz, 10-Series, Champlain, NY) (Fig. 2b). A benefit of this construction is the absence of solid planar boundaries, allowing components to protrude in any direction, unencumbered by immovable apparatus walls. The extruded aluminum was used to build integrated mounts for the imaging system, coil mounts (see Sec. II.E), and other minor components. The frame includes eight low-durometer vibration-dampening feet (McMaster Carr, part# 60855K72, Chicago, IL) that enhance image quality. A removeable sub-frame with castors slides under the bottom of the bench and allows the system to be easily moved. Amplifiers, computer, monitor, and auxiliary components were housed separately on a mobile equipment rack (Samson, SRK-21, Hicksville, NY) to isolate the µwheel driving and imaging components from heat and vibration (Fig. 2a).

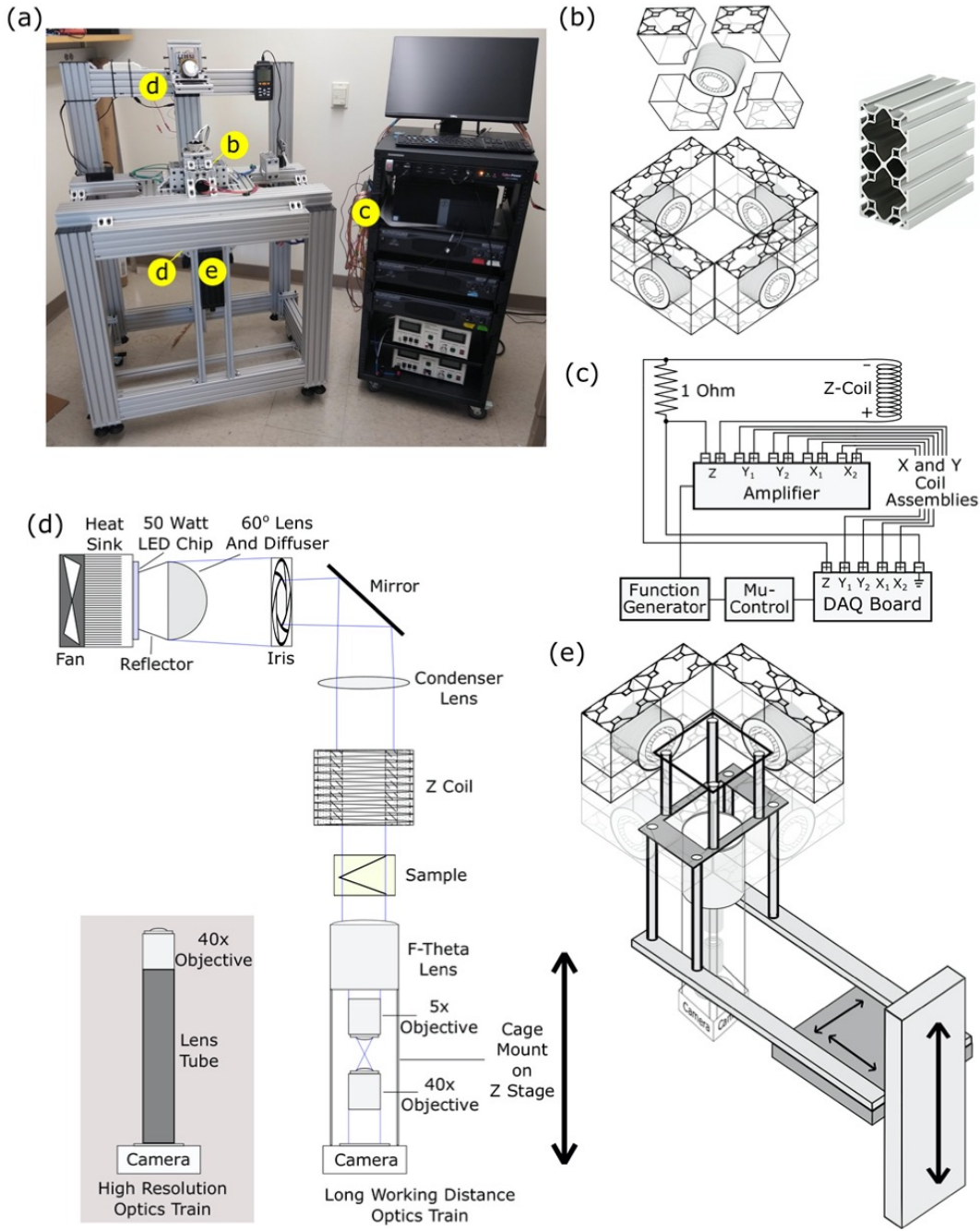

**FIG. 2.** (a) µWheel manipulation bench and equipment rack. (b) Schematic of coil holders with exploded view of the z-coil holder and a cross-section of the extruded aluminum framing, (c) circuit diagram for current measurement system, (d) schematic of imaging components, including the short focal distance, high resolution optics train in the inset, (e) schematic of translational stage.

**C. Magnetic Field Generation: Hardware**

A rotating magnetic field assembles superparamagnetic particles (e.g. 4.5 µm, Life Technologies, Dynabeads M-450 Epoxy, Carlsbad, CA) into µwheels and drives them along surfaces (Fig. 1). The rotating magnetic field is created with five coils (Fig. 2b). Two pairs of coils with custom-made cores were mounted in the x-y plane and a fifth coil was oriented orthogonally, in the z-direction. Assembly and rotation in the x-y plane were achieved by two sine waves with a 90° phase shift controlled by home built software (Sec. II.H), conditioned with an analog output card (NI-9263, National Instruments, Austin, TX), amplified (EP2000, Behringer, Willich, Germany), and sent to each pair of coils. The fifth coil is supplied with its own phase-shifted AC signal by an additional amplifier and independent sine wave. With the five coil system, µwheels can be set to any camber angle, $\theta_c$, from 0° to 90 using a 3D rotating magnetic field.

A 1 Ω resistor in series with each coil (Fig. 2c) was used to monitor the current and associated voltage drop by a DAQ board (NI-USB-6009, National Instruments, Austin, TX). Each amplifier outputs to two channels and can supply 650 $W_{RMS}$ of power per 2 Ω channel. Each coil assembly, with winding layers wired in parallel (Sec. II.D), has a resistance of 2.2 Ω.

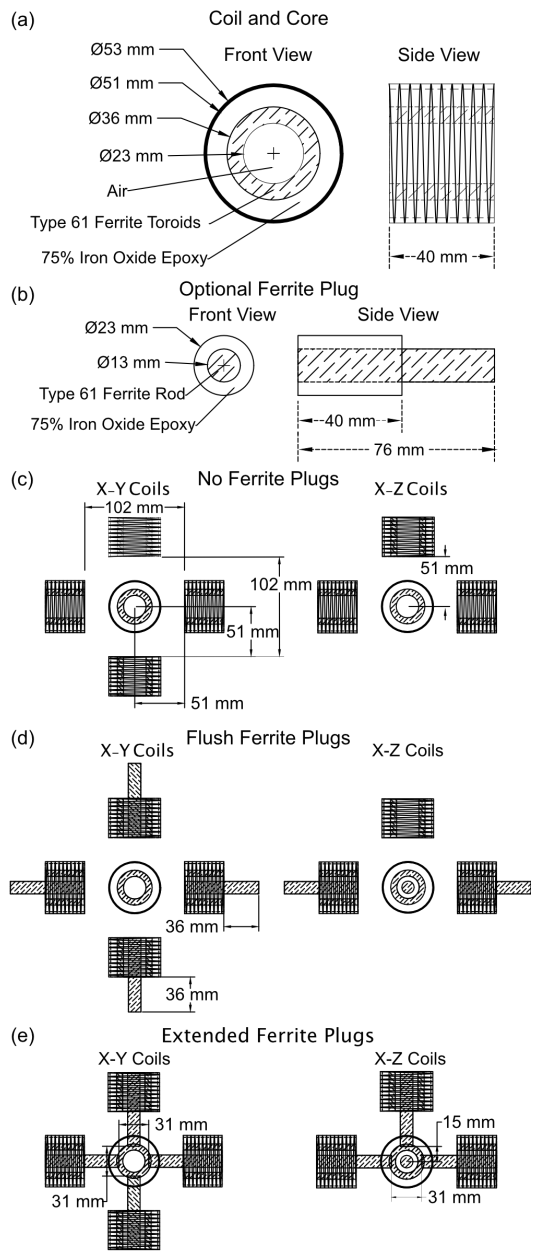

**FIG. 3.** Schematic of coils and cores. (a) Typical coil and core used for x, y, and z coils, (b) ferrite plug which can be inserted within the ferrite toroid void space within each coil core, (c) spacing of coils around the test section with no ferrite plugs, (d) coil spacing around the test section with flush mounted ferrite plugs, (e) coil spacing with extended ferrite plugs, making a smaller test section.

Each coil circuit, including the 1 Ω resistor and associated wiring, has a resistance of 3.3 Ω. With this loading, the amplifiers can provide 43 $V_{RMS}$ at a current of 13 $A_{RMS}$ per channel; however, the DAQ card has a maximum input voltage of 8.5 $V_{RMS}$, limiting the maximum current supplied to each coil to 8.5 $A_{RMS}$.

### D. Electromagnetic Coils and Cores

Voice coils used for audio speakers (Springfield Speaker, 2" Kicker Voice Coils 4 Ω, Springfield, MO) were used to generate the magnetic fields. Each coil is 53 mm in diameter by 40 mm long and is comprised of four winding layers, with 84 windings per layer, giving 336 total windings of 26 awg round aluminum wire. Ferrous cores were incorporated into the coils to increase the magnetic field amplitude while preserving optical access through the center (Fig. 3a). Three type-61 ferrite toroids (AnaTek Instruments, Type 61 Ferrite Toroid, Santa Clara, CA) were stacked to form a 38 mm long cylinder with an inner diameter of 28 mm and outer diameter of 36 mm. The stack of toroids was centered within the coil and a mixture of 75% iron oxide (by volume) ferrous epoxy material was poured between the coil former and the toroids. The ferrous epoxy was made by mixing iron oxide powder (Alpha Chemicals, Natural Black Iron Oxide, Missouri) with clear epoxy resin (East Coast Resin, Crystal Clear Epoxy Resin).

For the x-y coils where optical access is not needed, optional ferrite plugs were designed to fit within the open area of the toroids (Fig. 3b). These plugs were made from ferrite rods (AnaTek Instruments, Type 61 Ferrite Rod, Santa Clara, CA), with 75% iron in epoxy ferrous material molded around the rods to make the geometry match that of the toroid void space. These plugs could be inserted in either a coil-face flush orientation to preserve the large test section volume (Fig. 3d) or with the ferrite rod protruding into the test section (Fig. 3e), creating a higher field strength for smaller samples.

**E. Coil Heat Management**

The magnetic field produced by a coil decreases as the coil temperature increases. If resistive heating of the coil becomes severe, the wire windings can melt, disrupting the circuit and causing coil failure. To lengthen the operating time and maintain steady-state within the magnetic field generation system, a coil cooling system was integrated within the apparatus design (Fig. 2b); 54 mm holes were bored through 100 mm lengths of the aluminum extrusion to house the magnetic coils. These sections were cut into four pieces, such that the 54 mm hole was in quarters. The four pieces could then be clamped around the coil, providing a solid mounting connection while also giving more direct contact between the heat producing coil and conductive coil holder (Fig. 2b).

**F. Imaging**

Two interchangeable optics trains were built to allow for visualization of different sample types. The long working distance train was designed to image large 3D samples using an f-theta lens (Fig. 2d). The short focal distance train was designed to capture high-resolution images on planar samples using a standard microscope objective (Fig. 2d). The long working distance optics train was designed to provide imaging at any point within a 10 cm cube. It incorporates an f-theta lens (Thorlabs, LSM05-BB, Newton, NJ) with an effective working distance of 11 cm, in conjunction with a pair of microscope objectives, one of which (Olympus, UPlan Apo 4x/0.16, Tokyo, Japan) is reversed to collimate the image coming through the other objective (Labromed, LW Ph Plan 40x/0.60, Los Angeles, CA). Images are captured with a CMOS camera with a 1280 pixel × 1024 pixel resolution (Thorlabs, DCC1645C, Newton, NJ) and logged using software (ThorCam version 3.3.1) supplied with the camera. To adjust focus, the camera and optical components are mounted to an extruded aluminum section coupled to a z-axis translational stage (Newport, M423, Irvine, CA) that is mounted to the apparatus frame. The interchangeable high resolution, short focal distance optics train for planar samples includes a 20X objective (Olympus, LUCPLanFLN 20x/0.45 RC2, Tokyo, Japan) mounted to one end of a lens tube (Thorlabs, SM1L40, Newton, NJ), which is coupled to the CMOS camera with cage plates (Thorlabs, CP02, Newton, NJ) and additional lens tubes. The working distance of this optics train is 7.8 mm. The cage plates are attached to brackets that can be easily installed on the same extruded aluminum section used by the long focal distance train, allowing focus via the z-axis translational stage.

Illumination is provided by a 50 watt LED chip (Chanzon, 1DGL-JC-50W-NW, Huaqiang, China) coupled to an aluminum heat sink (TX, H&PC-73411, FXT Technologies, Shenzhen, China). To dissipate heat emanating from the light source, the heat sink is mounted to the aluminum apparatus frame and fit with a small fan (Fig. 2d). Light is sent through a diffuser, focusing lens, an iris for intensity adjustment, reflected by a mirror, and then refocused to intercept the imaging path. A custom sample holder was constructed from a vertical actuator (Link CNC, 4080U, Jinan, China) affixed to an x-y translational stage (AmScope, GT100 X-Y Gliding Table, Irvine, CA) which holds a specifically built riser and glass sample holder. The geometry of the riser and sample holder were designed to accommodate the geometry of both imaging and magnetic field generation components while simultaneously allowing translation of the sample (Fig. 2e). The lower riser section was designed to straddle the long working distance optics train, allowing 5 cm of stage movement in all three-dimensions. The top riser was designed to fit inside the coils, allowing 5 cm of sample holder movement in all three-dimensions without contacting coils or coil holders. To eliminate vibrations due to the rotating magnetic field, all stage components proximal to the magnetic fields are made of non-magnetic materials including aluminum (riser legs), glass (sample holder), and nylon (posts to affix glass sample holder to riser legs).

**G. Test Sections**

For μwheel driving tests on planar surfaces, a polydimethylsiloxane (PDMS; Sylgard 184, Krayden, Denver, CO) test section was fabricated [21]. The PDMS base and catalyst were thoroughly mixed at a 10:1 ratio. This mixture was poured into a rectangular, 25 mm x 30 mm x 3 mm mold, producing a 3 mm thick sheet of PDMS. The mold was then placed in a vacuum chamber for several hours to remove any entrapped air in the PDMS, then moved to an 80 ˚C oven overnight to fully cure. Once cured, the PDMS was removed from the mold and three 3 mm diameter holes were punched (Integra, Miltex 3332P/25, Princeton, NJ) through the sheet of PDMS to create three evenly spaced wells and placed on a clean glass slide. Each well was filled with a 0.1% Dynabead® suspension of 4 x $10^5$ particles/mL in deionized (DI) water with 1.4% sodium dodecyl sulfate (SDS, CAS 151-21-3, Fisher Bioreagents, Pittsburg, PA). A cover slip was then placed over the wells.

For μwheel driving tests in 3D, a helical tube with elevation change in the z-direction was fabricated using a spring with a circular 2 mm diameter cross section, 21 mm coil diameter, and a 9° pitch. The spring was secured inside a rectangular Petri dish, creating a mold with planar sides. PDMS was poured into the mold, filling the volume surrounding the spring. To track μwheels within the test section, alternating colors of candle wax dye (Candlewic, Doylestown, PA) were injected into the PDMS, providing visual cues for location within the helical channel. The mold was degassed and cured using the same procedure as the planar test section. Once cured, the spring and Petri dish molds were removed from the PDMS and the test section was trimmed to remove excess material. The helical tube was saturated with 1.4% SDS in DI water from the bottom elevation up using a hypodermic needle. Then, a 50 μL bolus of 0.5% Dynabead suspension of 2 x $10^6$ particles/mL in DI water with 1.4% sodium dodecyl sulfate was injected into the bottom elevation of the tube. Finally, the top elevation of the channel was plugged with a modified pipette tip.

**H. Magnetic Field Generation: Software**

A multithreaded open-source Python application, MuControl [15], was written for the generation, control, and monitoring of signals in the coils required to drive μwheels. Visualizations and live plots within the application allow for confirmation of the status of the magnetic field during use. Calibration tools enable simple output waveforms for testing and measurement of each individual coil set. MuControl is packaged and compiled into a self-contained executable file, allowing for compatibility with most computer hardware, eliminating the need for up to date Python distributions or coding experience. The user interface was designed using Python PyQt5 [16] and pyqtgraph [17] packages and provides the user with a live signal plot, a visualization of the shape of the magnetic field, and controls for both rolling and calibration signals (Fig. 4).

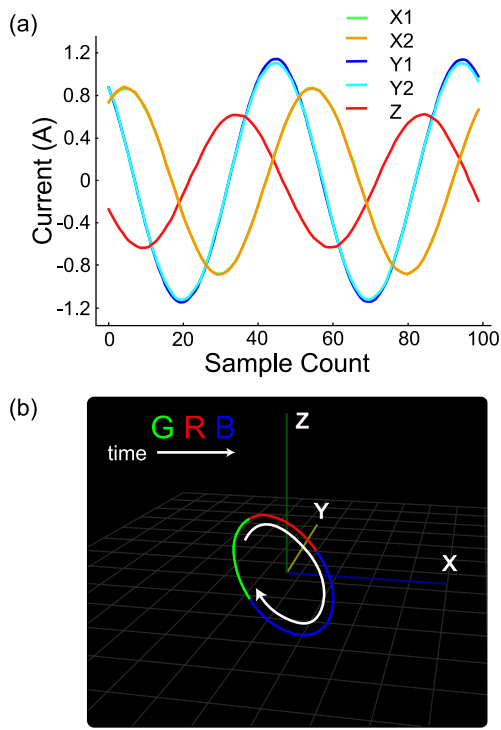

**FIG. 4.** MuControl user interface, consisting of (a) a real-time signal plot and (b) 3D parametric view of the output coil voltages.

The application interfaces with an analog output card (National Instruments, NI-9263) and an analog input data acquisition card (National Instruments, NI-USB-6009) using the Python nidaqmx library [18]. This program is written based on the software created in our earlier work [12] with improvements such as open-source application, self-contained executable file and a user interface allowing easier setting of the input parameters. Concurrent processing threads monitor these tasks and forward data to the user interface. Given the desired rolling direction, frequency, magnitude, $\theta_c$ supplied by the user, three AC signals are calculated and output. The magnetic fields from the three coil axes superimpose in the working volume, forming the desired three-dimensional rotating magnetic field to control and direct the μwheels. The amplitude, frequency, and phase of these waves can be confirmed in the live signal plot. The signal parameters are controlled in the user interface using a keyboard, gamepad, or joystick, functionality implemented using the XInput-Python library [19]. The default signal refresh rate of 10 Hz allows for manipulation of µwheels in real-time with no observable input lag.

A settings window (Fig. S1) with common parameters is available for modification of signal generation and read rate, output card type, and number of channels to complement the modular design of the entire system. The application does not require recompiling for changes in equipment type or number of coils, a feature that allows MuControl to work for many different coil systems. The live plotting frequency can be adjusted to decrease the number of lines written to the screen per second if high frequency monitoring of the signal is not required. A Python plotting library, PyQtGraph, is used to decrease the impact on the target CPU [16]. Calculation, manipulation, and storage of signals are performed using the NumPy Python package [17]. It is packaged as a self-contained executable using the fbs library [20].

**I. Quantification methods**

*1. Magnetic field generation*

Magnetic fields were measured for the five coil system for toroid cores with no ferrite plugs (Fig. 3c), for toroid-cores with flush mounted ferrite plugs in the x-y coils (Fig. 3d), and for toroid-cores with the extended ferrite plugs in the x-y coils (Fig. 3e). Coils were powered using an AC current with a peak magnitude of 4 A (4 $A_p$) and a frequency of 40 Hz. Field measurements were made with a Gaussmeter (Latnex, MF-30K, Ontario, Canada), with the probe attached to a precision guide, making measurements at 1 cm increments over a 10 cm × 10 cm grid for the x-y plane and over a 10 cm × 5 cm grid for the x-z plane. Centerline measurements of field strength were taken in the x-direction at z = 0 at 1 cm increments for air-cored and toroid-core coils with and without ferrite plugs in the configurations shown in Figs.

3c-d. Centerline measurements of field strength were taken in the x-direction at z = 0 at 0.5 cm increments for toroid-core coils with extended ferrite plugs (Fig. 3e).

## *2. Heat generation*

Heat generation tests were performed at an AC current of 4 $A_P$ and frequency of 40 Hz to determine the temperature of the windings of a single coil. Measurements were taken for air-cored coils with no holder, air-cored coils with the heat-dissipating holder, and toroid-core coils with no ferrite plug with the heat-dissipating holder. Temperature measurements were also taken at the center of the test section for the five-coil system to measure sample heating. System measurements were taken for toroid-core coils with either x-y coils only or x-z coils only housed in heat-dissipating coil holders and air-cored x-y coils with no heat-dissipating coil holders. Temperature was measured using a thermistor (Cole-Parmer, Digi-sense 20250-93, Vernon Hills, IL) placed in direct contact with the top of the coil for single coil tests and in the center of the test section for five-coil system tests. To allow coils to reach temperature steady state, measurement durations were in excess of 1 hr.

## *3. μWheel driving*

To test μwheel translation on planar surfaces, hexagonal μwheels consisting to seven particles were driven on a planar surface at frequencies of 10 Hz, 20 Hz, and 30 Hz, at $\theta_c = 30°$. Images were captured using the short focal distance optics train. To test μwheel driving in a 3D environment, a swarm of μwheels were driven through a helical channel at a frequency of 40 Hz at $\theta_c = 20°$. Images were captured using the long working distance optics train. In both experiments, the toroid-core coils with no ferrite plugs were used (Fig. 3c). A current of 4 $A_P$ was used for coils in the x-y plane and to match field strength with the single z-coil, 8 $A_P$ for the z-plane.

For the planar surface, µwheel velocity was calculated as the average of two measurements of wheel translation distance over time, using the same 7-mer wheel traveling in opposite directions for each measurement. For all measurements used to study the effect of varied frequencies, the same 7-mer µwheel was driven over an identical planar surface. For the helical channel, images were taken of each section of the channel to record particle concentration. A bolus of particles was first injected into the bottom of a spiral channel, the field was turned on, and wheels were allowed to drive up the spiral for 3.7 min. The field was then removed and channel images were captured. An identical magnetic field was then applied for another 3.3 min with no additional user input. After removing the field and acquiring images, the field was reapplied at $\theta_c = 20°$ with a single heading-correction of 90° to accommodate a change in direction dictated by the helical channel geometry. µWheels were then allowed to translate for another 4.5 min.

### J. Simulations

#### *1. Finite element method simulations of magnetic fields and heat transfer*

The magnetic field and heat generation were simulated using COMSOL Multiphysics 5.4. A 3D geometry was created wherein five coils were defined within a spherical domain of radius 0.5 m and housed within aluminum scaffolding according to the design specifications and dimensions in Sec. II.C (Fig. 5). The built-in material properties for 26 AWG aluminum were used for the coils, 6063 aluminum alloy for the coil holders, and air for the spherical domain. Custom material definitions were created for the iron oxide epoxy and ferrite (see Sec. II.J.2).

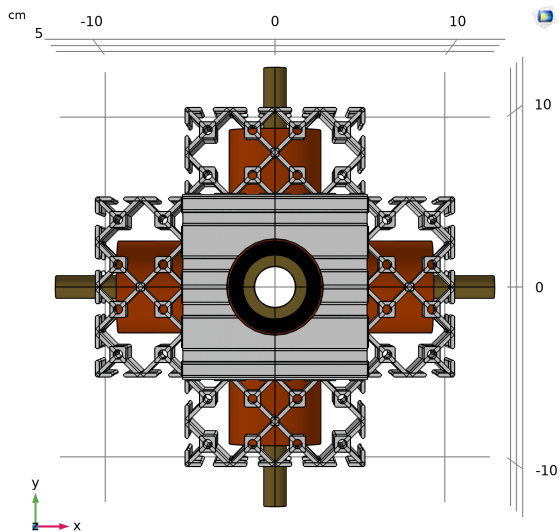

**FIG. 5**. Top-down view of COMSOL geometry with optional plugs inserted in x and y coils in the "flush" position. Materials are aluminum (silver), copper (copper), ferrite (gold), and iron oxide epoxy (black).

The COMSOL magnetic field (.mf) module was used to estimate the magnetic flux as a function of position using an input of 4 A of current at 40 Hz for each coil to match experimental conditions. For all coils, the direction of current flow was defined around the inner edge. The electrical currents (.ec) and heat transfer in solids (.ht) modules were coupled to calculate the joule heating for each coil. Heat transfer simulations were accomplished using the heat transfer in solids and fluids module (.htsf) by treating the coil surfaces as heat sources with an output equal to their Joule heating. In this, all solid-solid and solid-fluid contacts are conductive boundaries. A Dirichlet boundary condition is used at the edge of the spherical air domain, where the temperature is fixed at 20-25 °C to match experimental room temperature. The size of the spherical domain did not significantly affect results at r > 0.15 m. For simulations with natural convection, a heat loss of $\dot{q} = h(T_c - T_a)$ was integrated for the coils, where $\dot{q}$ is the heat dissipated per unit area, $h$ the natural convection heat transfer coefficient (12.45 W/m$^2$), $T_c$ the temperature of the coil surface and $T_a$ the temperature of the surrounding air.

A free tetrahedral mesh was defined with a maximum element size of 0.1 cm, a minimum element size of 0.02 cm, a maximum element growth rate of 1.6, a curvature factor of 0.7, and a narrow resolution of 0.4. Convergence tolerance was $10^{-4}$. The residual root mean square error was of order $10^{-31}$, and parameter imbalances across the domain were <$10^{-6}$ % after double the required iterations to converge, indicating mesh independence of the study. The solution used a domain consisting of 1.5 × 10⁶ elements and did not significantly change when using a finer (4.3 × 10⁶ elements) or coarser (1.5 × 10⁵ elements) mesh (Fig. S2).

TABLE 1: Custom material parameters for COMSOL model

|  | μ | σ | ε (W/m*K) | ρ (kg/m³) | $C_P$ (J/kg*K) |
|---|---|---|---|---|---|
| **Epoxy** | 1.7 | 0.5 | 10.2 | 3960 | 1000 |
| **Ferrite** | 125 | 0.01 | 4 | 5000 | 750 |

### *2. Physical properties of epoxy and ferrite*

To compute the electromagnetic and heat transfer physics for the geometry, five parameters were required for each material: relative permeability μ, relative permittivity σ, density ρ, heat capacity $C_P$, and thermal conductivity k (Table 1). Relative permeability and relative permittivity were measured using a magnetometer (MPMS3, Quantum Design, San Diego, CA) at 25 °C from -1 to 1 T. The density of ferrite is reported to be ~5000 kg/m³ by the manufacturer (AnaTek, Santa Clara, CA) and the density of iron oxide epoxy was approximated using a weighted average of the densities of iron oxide and epoxy. Heat capacity was measured using calorimetry where a sample of ferrite or iron oxide epoxy was massed and suspended by sewing thread within a water bath at 90 °C for 2 hr. Thermal conductivity, ε, for the iron oxide epoxy was determined by measuring the temperature difference between the top and bottom of an insulated iron oxide epoxy cylinder with heat supplied at constant power to the bottom of the cylinder.

## III. RESULTS

### A. Magnetic field generation

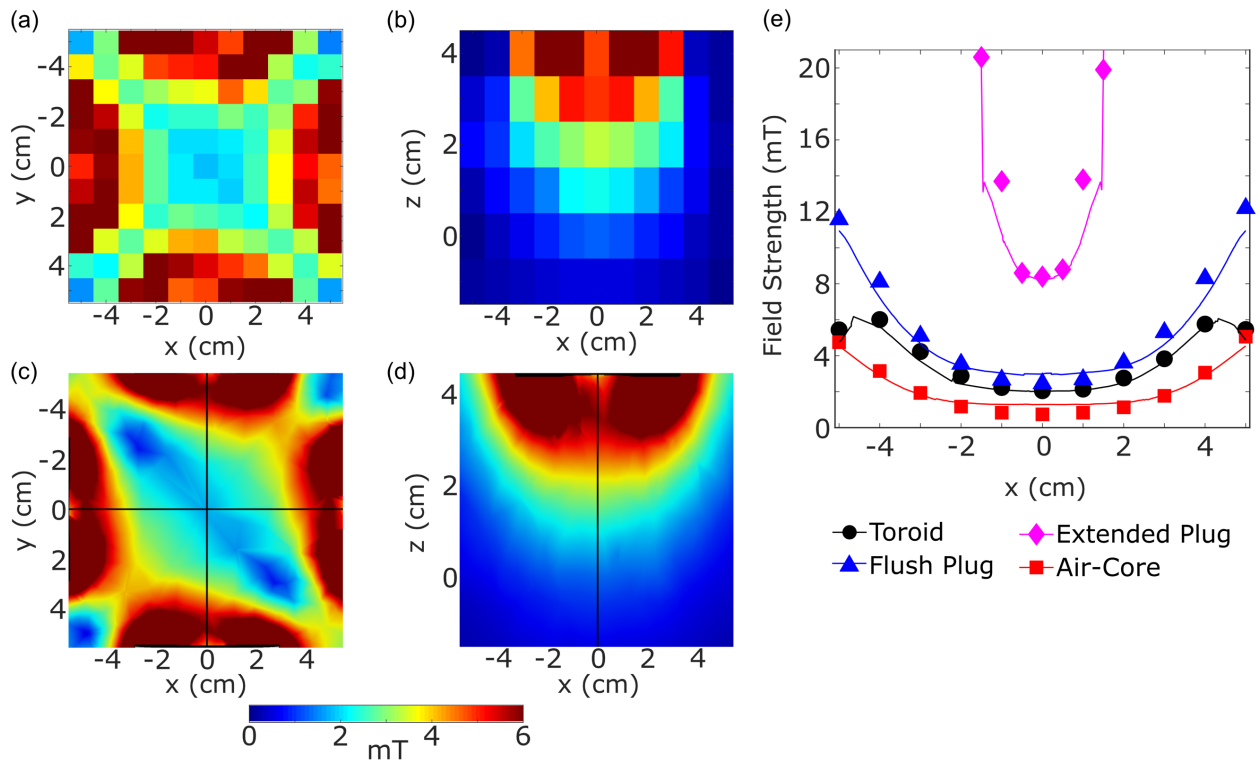

**FIG. 6.** Measured and simulated magnetic field magnitudes for (a,c) x-y plane coils and (b,d) z-plane field measurements for toroid-cored coils. Data in (a, c) were collected or (c,d) simulated using 4 $A_p$ of AC current at 40 Hz. e) Measurements (symbols) and simulations (lines) at y = 0 for toroid-cored coils, toroid-cored coils with flush mounted ferrite plug, toroid-cored coils with extended ferrite plug, and air-cored coils.

Two-dimensional representations of the magnetic field in the x-y plane with toroid-cores and no ferrite plugs indicate that the field is symmetrical, with field magnitudes near 2 mT at the center point and increasing field strength with proximity to the coil (Fig. 6). An exception exists at regions very near the center of the coil where the void within the toroid core creates a lower field magnitude (Fig. 6a). The z-coil field representation also shows a symmetrical field, with approximately 1 mT at the sample (Fig. 6b) with an applied current of 4 $A_P$. The lower field strength compared to the x-y plane is due to the use of a single coil rather than a coil pair along the z axis. COMSOL simulation results are in agreement with experimental measurements (Fig. 6d-e). The x-y plane measurements and simulations both exhibit slight asymmetry across quadrants. Asymmetries between the Northwest/Southeast vs. Northeast/Southwest quadrant pairs are attributable to the magnetic interactions between coils and the direction of current flow within those coils.

Centerline measurements show how different coil configurations can be used to enhance field strength (Fig. 6e). Toroid-core coils with flush mounted ferrite plugs gives the largest fields compared to toroid-core coils with no ferrite plug and air-cored coils. Toroid-core coils with extended plugs yield the overall greatest field strength in a more focal volume owing to the ability of the plugs to act as a more efficient conduit for magnetic force than air.

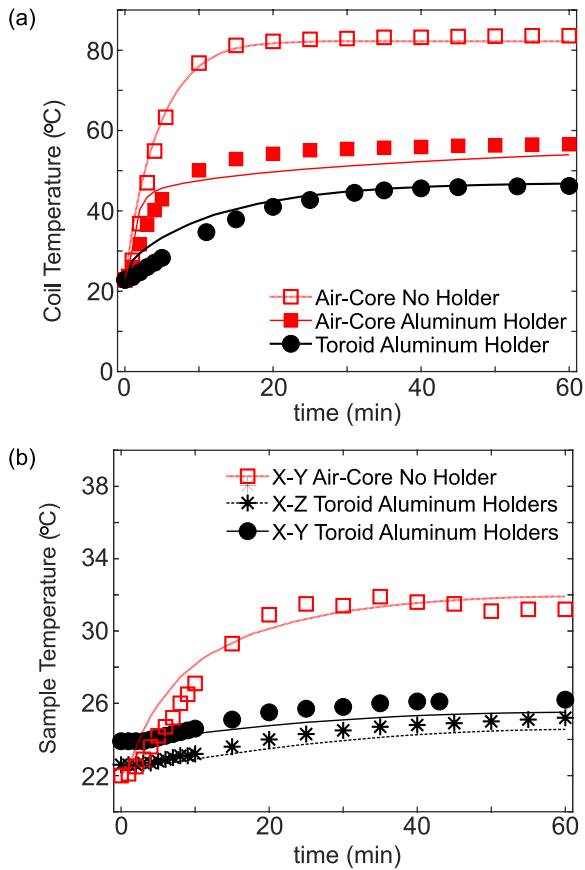

**FIG. 7.** Coil heat generation with time. (a) Data (symbols) and simulations (lines) of temperature of a single coil for an air-cored coil with no coil holder, air-cored coil with aluminum holder, and a coil with the iron oxide epoxy and type 61 toroid ferrite core housed in an aluminum holder. (b) Data (symbols) and simulations (lines) of temperature at sample for an air-cored x-y coils with no coil holders, and x-y and x-z ferrite-cored coils with aluminum coil holders. All data collected at 4 $A_p$ of AC current at 40 Hz.

### B. Heat generation results

Securing coils in aluminum clamps showed improved heat dissipation compared to clamps where most of the coil surface area is exposed to air (Fig. 7). For single coil tests, the air-cored coil with no heat-dissipation reached a maximum temperature of 83.6 ˚C while those in aluminum clamps reached 56.6 ˚C. The toroid-core coil had an even lower temperature of 46.4 °C due to the higher conductivity of the ferrite compared to air. All single coil tests were performed at 4 $A_p$ of AC current at 40 Hz.

To measure the temperature at the sample, either x-y pairs or x-z pairs were tested with and without the heat dissipating aluminum coil holders (Fig. 7b). Toroid-core coils with heat-dissipating holders stayed within 3 °C of the ambient temperature at the sample. Air-core coils without heat-dissipating clamps showed a 9.9 °C rise in temperature above ambient at the sample.

COMSOL simulations show good agreement with experimental results. For coil temperatures, simulations are within 3 °C of experimental measurements at steady state (Fig. 7a). Coil temperature can be used as a predictor for coil failure. Due to impurities in the aluminum wire or irregularities in electrical connections, failure temperature varies. Consequently, simulations that are within a few degrees of the actual temperature serve as a useful model for experimental design. For sample temperatures, simulations are within 1 °C of experimental measurements at steady state (Fig. 7b). Sample temperature is important for biological samples. In this case, simulations that are within a degree of actual temperature serve as a useful model for experimental design.

## C. μWheel driving

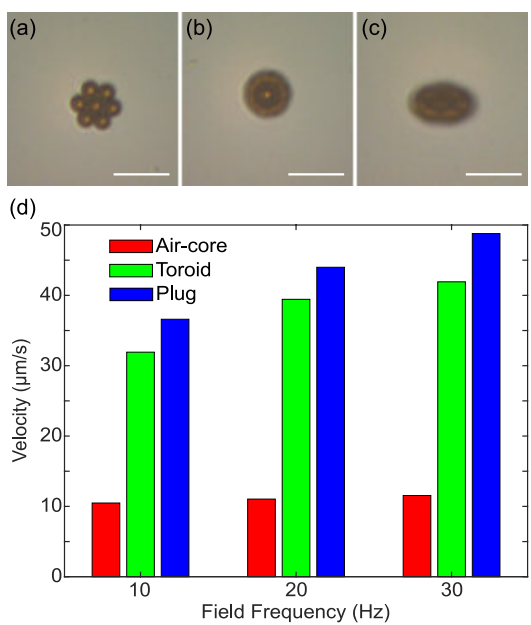

**FIG. 8.** μWheel driving on a planar surface. Images of a μwheel consisting of seven particles with a magnetic field turned off (a), in a x-y magnetic field causing it to spin (b), in x-y-z with $\theta_c$ = 30º. Scale bar = 20 μm. (d) Velocity as a function of field frequency with $\theta_c$ = 30º with magnetic fields generated with air-cored coils, toroid-core coils, and toroid-core coils with the ferrite plug.

Fig. 8 shows the velocity of hexagonal μwheels consisting of seven particles driven on a planar surface. In agreement with field strength measurements, toroid-core coils with ferrite plugs led to the fastest μwheel speeds followed by the toroid-core coils without ferrite plugs, and then air-cored coils. Similar results were found for various frequencies between 5 Hz and 40 Hz and for 30˚ < $\theta_c$ < 90˚ (Fig. S3).

μWheel driving was tested in a 3D helical tube where a bolus of particles was injected into the bottom of the tube (Fig. 9a-b). The rotating magnetic field was then applied at a current of 4 $A_P$ for coils in the x-y plane and 8 $A_P$ for the z-plane at 40 Hz. μWheels were visualized with the long working distance optics train (Fig. 9c). During translation, μwheel sizes ranged from single monomers to large aggregates of hundreds of particles. Aggregate sizes tend to increase with time as μwheels collide and combine with each other. Fig. 9d shows the distribution of μwheels at various times as the swarm translated upwards and along the tube, with wheels traveling 37 mm around the helix and reaching a vertical distance of 5.7 mm. Note that, due to the curvature of the channel cross-section, only particles in the center portion of the channel are visible. Therefore, the data presented in Fig. 9d is normalized as $N/N_{VIS}$, where $N$ is the number of particles within the bin and $N_{VIS}$ is the total number of particles visible throughout the channel at the specified time.

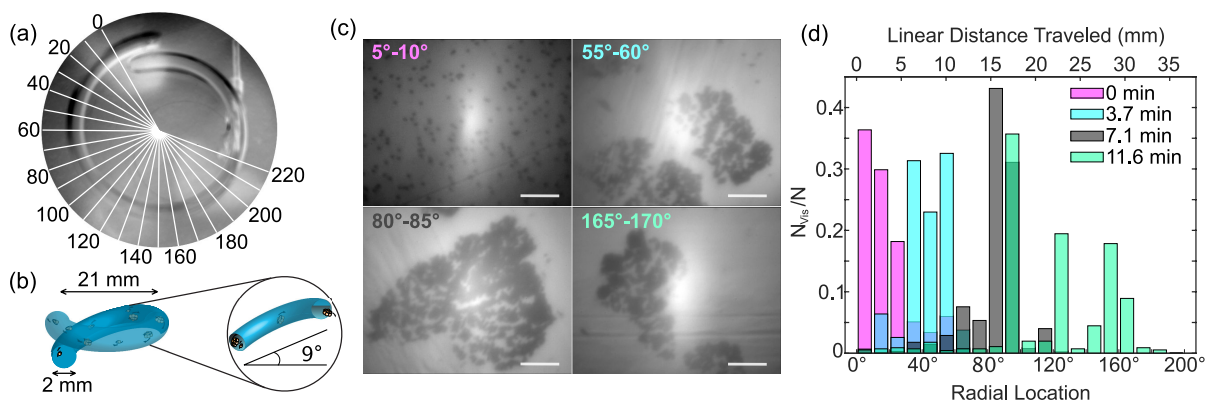

**FIG. 9.** µWheel driving through a helical channel. (a) Top view of the PDMS helical channel and overlay of radial locations in degrees. (b) Side view helical channel schematic showing the circular channel with dimensions and pitch. (c) Representative images of µwheels in the 5º-10º, 55º-60º, 80º-85º, and 165º-170º arcs of the helix. Scale bar = 50 µm. (c) Distribution of particles within the helical channel at different locations and times normalized by the total number of visible particles within the channel at each time.

## IV. SUMMARY

Here we present a scaled up design for a microbot manipulation apparatus allowing acceptance of larger, 3D samples than previous studies [12-14]. With this design, µwheels can be assembled, controlled, and imaged in 2D and 3D networks in real-time. This apparatus is capable of minimizing coil heat generation, with sample temperatures residing within a few degrees Celsius of ambient for continuous experimental run times in excess of one hour. Rotating magnetic fields on the order of a few milli-Tesla are capable of assembling and driving wheels in centimeter scale 2D and 3D samples. This low-cost apparatus, built from off-the-shelf components can be used to investigate µwheel use in a variety of applications including drug delivery and cargo transport .

## ACKNOWLEDGMENTS
22

The authors acknowledge support from the National Institutes of Health under grants R21AI138214, R01NS102465, and T32HL072738 (E.J.R.). D.D. was supported by an American Heart Association Predoctoral Fellowship Award 18PRE34070076.

**REFERENCES**

1. Jeon, S., et al., *Magnetically actuated microrobots as a platform for stem cell transplantation.* Science Robotics, 2019. **4**(30): p. eaav4317.
2. Mellal, L., et al. *Magnetic microbot design framework for antiangiogenic tumor therapy*. in *2015 IEEE/RSJ International Conference on Intelligent Robots and Systems (IROS)*. 2015. IEEE.
3. Yang, T., et al., *Magnetic microlassos for reversible cargo capture, transport, and release.* Langmuir, 2017. **33**(23): p. 5932-5937.
4. Guasto, J.S., R. Rusconi, and R. Stocker, *Fluid mechanics of planktonic microorganisms.* Annual Review of Fluid Mechanics, 2012. **44**: p. 373-400.
5. Abbott, J.J., et al., *Robotics in the Small, Part I: Microbotics.* IEEE Robotics & Automation Magazine, 2007. **14**(2): p. 92-103.
6. Chen, X.-Z., et al., *Recent developments in magnetically driven micro- and nanorobots.* Applied Materials Today, 2017. **9**: p. 37-48.
7. Hwang, G., et al., *Note: On-chip multifunctional fluorescent-magnetic Janus helical microswimmers.* Review of Scientific Instruments, 2016. **87**(3): p. 036104.
8. Zhang, L., et al., *Controlled propulsion and cargo transport of rotating nickel nanowires near a patterned solid surface.* ACS nano, 2010. **4**(10): p. 6228-6234.
9. Gong, D., et al., *Controlled propulsion of wheel-shape flaky microswimmers under rotating magnetic fields.* Applied Physics Letters, 2019. **114**(12): p. 123701.
10. Kummer, M.P., et al., *OctoMag: An Electromagnetic System for 5-DOF Wireless Micromanipulation.* IEEE Transactions on Robotics, 2010. **26**(6): p. 1006-1017.
11. Jiang, C., et al., *Electromagnetic tweezers with independent force and torque control.* Review of Scientific Instruments, 2016. **87**(8): p. 084304.
12. Tasci, T.O., et al., *Surface-enabled propulsion and control of colloidal microwheels.* Nature Communications, 2016. **7**: p. 10225.
13. Disharoon, D., K.B. Neeves, and D.W. Marr, *AC/DC Magnetic Fields for Enhanced Translation of Colloidal Microwheels.* Langmuir, 2019.
14. Tasci, T.O., et al., *Enhanced Fibrinolysis with Magnetically Powered Colloidal Microwheels.* Small, 2017. **13**(36): p. 1700954.
15. Zimmermann, C.J., *czimm79/MuControl-release*. 2019: GitHub.
16. *What is PyQt?* 2019, Riverbank Software.
17. *PyQtGraph*, in *Scientific Graphics and GUI Library for Python*. 2019.
18. *ni/nidaqmx-python*. 2019, National Instruments.
19. *Zuzu-Typ/XInput-Python*. 2019: GitHub.
20. Herrmann, M., *mherrmann/fbs*. 2019: GitHub
21. Duffy, D.C., et al., *Rapid prototyping of microfluidic systems in poly (dimethylsiloxane).* Analytical chemistry, 1998. **70**(23): p. 4974-4984.

# SUPPORTING INFORMATION

**Table S1. Apparatus parts and supplies**

| Part Description | Part Designation | Supplier |
|---|---|---|
| Extruded aluminum | 10 series extrusion and components | Tnutz, Champlain, NY |
| Vibration dampening feet | part# 60855K72 | McMaster Carr, Chicago, IL |
| Equipment rack | SRK-21 | Samson, Hicksville, NY |
| Superparamagnetic beads | Dynabeads M-450 Epoxy | Life Technologies, Carlsbad, CA |
| Analog output card | NI-9263 | National Instruments, Austin, TX |
| DAQ board | NI-USB-6009 | National Instruments, Austin, TX |
| Amplifiers (x3) | EP2000 | Behringer, Willich, Germany |
| Voice coils (x5) | 2" Kicker Voice Coils 4 Ω | Springfield Speaker, Springfield, MO |
| Type 61 ferrite toroids (x15) | Type 61 Ferrite Toroid | AnaTek Instruments, Santa Clara, CA |
| Type 61 ferrite rods (x4) | Type 61 Ferrite Rod | AnaTek Instruments, Santa Clara, CA |
| Epoxy resin | Crystal Clear Epoxy Resin | East Coast Resin |
| Iron oxide powder | Natural Black Iron Oxide | Alpha Chemicals, Missouri |
| F-theta lens | LSM05-BB | Thorlabs, Newton, NJ |
| 4x objective lens | UPlan Apo 4x/0.16 | Olympus, Tokyo, Japan |
| 40x objective lens | LW Ph Plan 40x/0.60 | Labromed, Los Angeles, CA |
| Camera | Thorcam color CMOS camera, part #DCC1645C | Thorlabs, Newton, NJ |
| Single axis translational stage | M423 | Newport, Irvine, CA |
| 20x objective lens | LUCPLanFLN 20x/0.45 RC2 | Olympus, Tokyo, Japan |
| Lens tube | SM1L40 | Thorlabs, Newton, NJ |
| Cage plates | CP02 | Thorlabs, Newton, NJ |
| 50W LED chip | 1DGL-JC-50W-NW | Chanzon, Huaqiang, China |
| Heat sink | H&PC-73411 | FXT Technologies, Shenzhen, China |
| Vertical Actuator | 4080U | Link CNC, Jinan, China |
| Two axis translational stage | GT100 X-Y Gliding Table | AmScope, Irvine, CA |
| PDMS | Sylgard 184 | Krayden, Denver, CO |
| Candle wax dye | Candle wax dye | Candlewic, Doylestown, PA |
| SDS | CAS 151-21-3 | Fisher Bioreagents, Pittsburg, PA |
| Gaussmeter | MF-30K | Latnex, Ontario, Canada |
| Thermistor | Digi-sense 20250-93 | Cole-Parmer, Vernon Hills, IL |

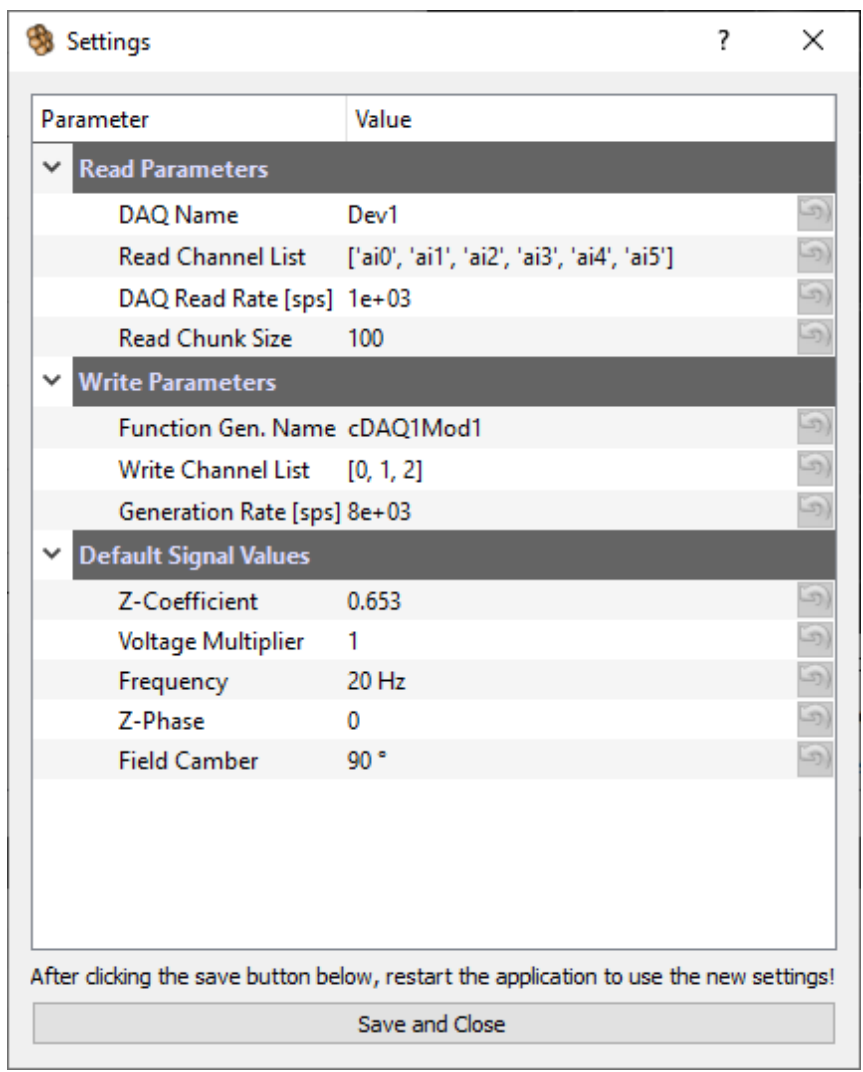

**Fig. S1.** MuControl settings window.

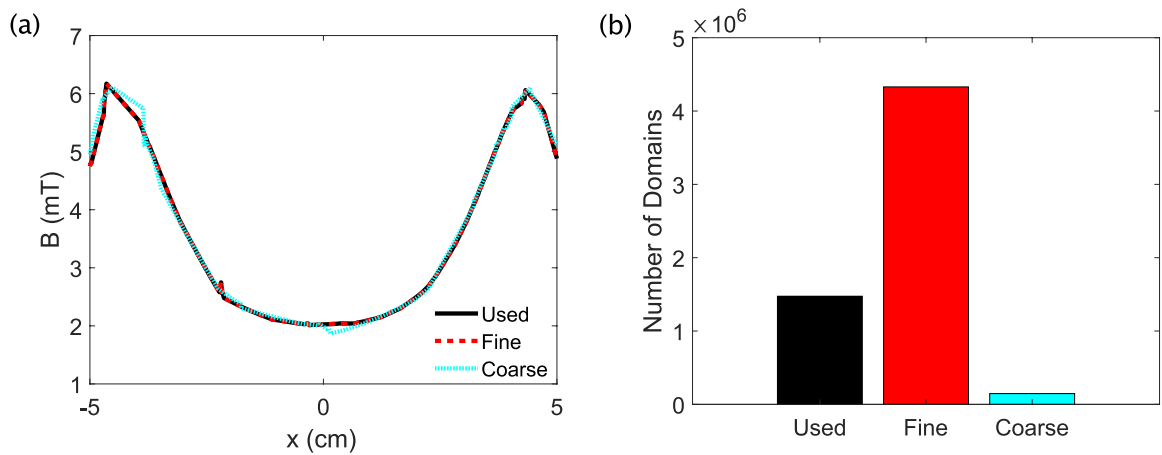

**FIG. S2**. (a) Simulated magnetic field profile along the x-axis for toroid-cored coils using 4 A of AC current at 40 Hz. We compare the solution for the mesh size used in our studies (black) to finer (red) and coarser (cyan) meshes. Our solution across the domain is on average <0.0002% different from the solution generated using the finer mesh and <1% different from that generated with the coarser mesh, indicating that our solution is mesh-independent. (b) Number of domains in each type of mesh.

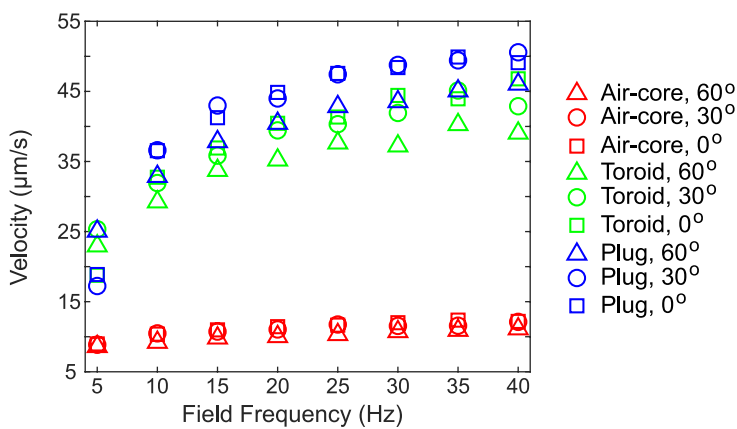

**FIG. S3.** Microwheel driving velocities on a planar surface at various field strengths, frequencies, and $\theta_c$.

27